\documentstyle[aps,epsf,prb]{revtex}
\input psfig.sty

\def \energy{\varepsilon}
\def \dwave{$d$-wave }
\def \bare{\chi_0}
\def \vertex{{\chi_{\rm corr}}}
\def \real{\chi_0'}
\def \reg{\chi_{\rm 0,reg}''}
\def \sing{\chi_{\rm 0,sing}''}
\def \upup{\chi^{\uparrow \uparrow}}
\def \downdown{\chi^{\downarrow \downarrow}}
\def \updown{\chi^{\uparrow \downarrow}}
\def \downup{\chi^{\downarrow \uparrow}}
\def \tildemu{\tilde{\mu}}
\def \order{{\cal O}}
\def \principal{{\cal P}}
\def \chia{\chi_{\rm A}}
\def \chib{\chi_{\rm B}}
\def \gam{\gamma}
\def \Q{{\bf Q}}
\def \k{{\bf k}}
\def \q{{\bf q}}
\def \p{{\bf p}}
\def \i{{\bf i}}

\def \x{{\bf x}}
\def \y{{\bf y}}

\begin{document}
\title{Susceptibility and vertex corrections for a square Fermi surface}
\date{December 3, 1997}
\author{D. Djajaputra and J. Ruvalds}
\address{Department of Physics, University of Virginia, Charlottesville, VA 22903}
\maketitle

\begin{abstract}
We investigate the response of an electron system which exhibits ideal nesting features.
Using the standard Matsubara formalism we derive analytic expressions for the imaginary
and real parts of the bare particle-hole susceptibility.
The imaginary part has sharp peaks whose maxima 
at the nesting momenta approximately scale with $(\omega/T).$ 
The peak lineshapes resemble neutron 
scattering data on chromium and some copper oxide superconductors. 
The real part of the bare susceptibility at the nesting vectors
diverges logarithmically at low temperatures. 
Analytic formulas for the first vertex correction to the  
susceptibility are derived for a Hubbard interaction,
and its momentum and temperature variations are calculated numerically. 
This term detracts substantially from the ordinary RPA terms for intermediate
values of the Coulomb repulsion. Exact cancellation of a certain class of diagrams 
at half filling is shown to result from particle-hole symmetry.  
We discuss the consequences of these results for spin fluctuation 
theories of high temperature superconductors and spin density wave instabilities.
\end{abstract}


\section{Motivation}

Nested Fermi surfaces with nearly parallel orbit segments
are traditionally associated with electronic instabilities. A peak
in the susceptibility at a nesting momentum   
can create a charge density wave
for strong electron-phonon coupling, or alternately induce a spin density
wave (SDW) when the Coulomb repulsion dominates.\cite{overhauser} 
These instabilities often appear in systems with reduced dimensionality
for electron dynamics like the quasi one-dimensional organic metals 
and layered structure materials.\cite{gruner} 

The present work on a square Fermi surface is motivated by the
discoveries of nesting phenomena in high temperature superconductors. 
In these materials, the anomalous normal state quasiparticle damping that
is linear in frequency and temperature can be explained by nesting, providing 
that electron-electron scattering is the primary damping source. 
The microscopic nested Fermi liquid\cite{nfl} (NFL) theory 
derives this unconventional damping from electron scattering across 
nested regions of Fermi surface that approximately satisfy the nesting condition   
$\varepsilon(\k + \Q)+\varepsilon(\k)=2 \mu,$ where $\Q$ is the
nesting vector. This implies scaling of the spin susceptibility at the nesting 
vector in frequency divided by temperature $(\omega/T)$ 
which has been observed by neutron scattering experiments. 
By contrast, the weak Fermi liquid damping that is the hallmark
of ordinary metals owes its quadratic frequency and temperature variation
to a susceptibility that is essentially independent of temperature.
Within a self-consistent scheme, the NFL theory has                      
provided a physical explanation of the optical conductivity, electronic Raman
spectrum, and many other normal state features of cuprate superconductors.\cite{review}
Fermi surface nesting has been discovered by photoemission\cite{shen} in 
${\rm Bi}_2 {\rm Sr}_2{\rm CaCu}_2{\rm O}_8$ and some other cuprate compounds.
 
Our present study is relevant to the spin fluctuation mechanism of \dwave
superconductivity which was originally examined thirty years
ago by Berk and Schrieffer.\cite{berk} The basic Fermi sphere yields extremely small
transition temperatures $T_c$ for $d$-waves, and thus this concept became dormant until
evidence for an unconventional mechanism in heavy fermion metals revived interest. The 
possibility of raising $T_c$ by a susceptibility enhancement treated in the random
phase approximation (RPA) was suggested by Scalapino {\it et al.}\cite{scalapino} and 
subsequently used by many groups in 2D tight-binding model calculations for cuprates.
Nesting enhances \dwave electron pairing in leading order exchange of antiferromagnetic
spin fluctuations.\cite{dpair} Schrieffer,\cite{schrieffer} 
however, has stressed the need to examine vertex corrections 
and other terms beyond the simple RPA. 
If the RPA series yields a large susceptibility enhancement, it is
reasonable to expect that self energy and vertex contributions become
correspondingly important.\cite{mahan}

Within the nesting approximation, 
Virosztek and Ruvalds\cite{vertexpreprint} have derived analytic 
expressions for the self energy and vertex corrections to the 
susceptibility at the nesting vector. These higher order contributions 
preserve the scaling of the susceptibility and they become comparable to the
RPA terms for intermediate values of the Coulomb repulsion. Thus the RPA method
by itself can be misleading, especially when the system is close  
to an SDW instability.  
 
In this paper we study these issues by using a 2D linearized dispersion model 
which gives a square Fermi surface at any filling. The linearization allows 
the imaginary and real parts of the noninteracting susceptibility to be obtained  
analytically and the perfect nesting allows us to examine the validity of the NFL theory.   

The linear dispersion model was first introduced by Mattis\cite{mattis} as a simplified version
of the 2D tight binding model. A similar model with square Fermi surface 
has also been studied by Luther.\cite{luther} 
Mattis\cite{mattis} and Hlubina\cite{hlubina} have claimed that the model is 
exactly solvable using bosonization and that the system is a 2D Luttinger liquid,
with spin-charge separation, rather than a Fermi liquid. Indeed, bosonization
is one of the powerful methods that can be used to solve the 1D version of this
model---which is a variant of the Tomonaga-Luttinger model.\cite{manybody} In higher dimensions, however, 
bosonization (at least in the simple version used by Mattis and Hlubina) 
is expected to give an incomplete picture of the system due to the presence 
of the particle-hole continuum.\cite{li} Haldane\cite{haldane} has pointed out that  
the distinctive feature of 1D fermion systems is the nonexistence of low-energy
particle-hole pairs with small momenta, whereas in higher dimensions this continuum always exists.    
This absence of excitation decay channels allows the 
response of a 1D system to be completely described in terms of
bosonic collective excitations. In the language of susceptibility, the imaginary part
of the spin and charge susceptibilities in 1D will consist only of delta function
peaks corresponding to these bosonic modes.\cite{mahan} In higher dimensions, 
we get a qualitatively different picture. In addition
to peaks from the collective excitations, there is also a regular contribution
to the susceptibility, corresponding to excitations of particle-hole pairs
which may also serve as a decay channel of the collective excitations if their
dispersions lie within the particle-hole continuum.  
This feature appears clearly in our analytic results for the susceptibility, and
may shed light to the current controversy concerning the existence of Luttinger 
liquids in dimensions higher than one.\cite{anderson}
 
We define the square model and our convention for units in 
\hbox{Sec. II.} In \hbox{Sec. III} we proceed to evaluate the imaginary and real parts of the
bare susceptibility and display the key features emanating from this square model 
in the noninteracting limit. Effects of electron-electron interaction are studied in
\hbox{Sec. IV} where we numerically compute the leading vertex correction to the susceptibility 
and compare it with the RPA term of the same order. In this section we also prove a 
particle-hole theorem that is relevant to a large class of diagrams near half filling.  
We discuss the results of our investigation in \hbox{Sec. V.}

\section{Square Fermi Surface Model}

We consider a modified Hubbard model on a square lattice defined by the Hamiltonian

\begin{equation}
H = H_0 + H_U = \sum_{\k, \sigma} (\energy(\k)-\mu ) \ c_{\k,\sigma}^+ c_{\k,\sigma}
+ U \sum_\i n_{\i, \uparrow} n_{\i, \downarrow},
\label{hamiltonian}
\end{equation}

\noindent with a noninteracting energy dispersion given by

\begin{equation}
\energy (\k)=v_F ( \ \left| k_x \right| +\left| k_y \right| \ ),
\end{equation}
for momenta in the first Brillouin zone $-\pi/a \leq k_x, k_y \leq \pi/a$.
In the coordinate space the noninteracting Hamiltonian has the form

\begin{equation}
H_0 = - \sum_{\i,\sigma} \sum_{m = -\infty}^\infty t(m) \ 
( \ c_{\i, \sigma}^\dagger c_{\i + \x_m, \sigma}^{\vphantom{\dagger}}
\ + \ c_{\i, \sigma}^\dagger c_{\i + \y_m, \sigma}^{\vphantom{\dagger}} \ ),
\end{equation}
with $\x_m \equiv (2m - 1, 0),$ $\y_m \equiv (0, 2m-1),$ and hopping amplitudes

\begin{equation}
t(m) = {2 v_F \over \pi a} \ {1 \over (2m-1)^2} .
\end{equation}
Note that the electrons are only allowed to hop from one sublattice to another, but not between
sites in the same sublattice.

In the rest of this paper we will use the following convention for units:
$v_F = \pi/a = k_B = 1,$ where $k_B$ is the Boltzmann constant. 
The unit of energy is therefore $v_F (\pi /a) = {\rm W}/2,$ where
W is the noninteracting bandwidth. If the bandwidth were 2 eV,
temperature $T = 0.03$ would correspond to 360 K.
The symbol $\mu$ will be reserved for the chemical potential. We also define a useful variable
$\gam \equiv 1/(2T) = \beta / 2.$ 

The Fermi surface for this model with $\mu = 0.9$ is the square shown in  
\hbox{Fig. \ref{square_fermi_surface}.}
The arrows specify the nesting vectors for this value of the chemical potential 
that will be used throughout this paper.

Assuming the velocity to be independent of momentum avoids the logarithmic divergence in the
density of states that characterize other square Fermi surface models (such as the 2D tight binding
band at half filling). This allows us to study the characteristic features due to nesting separately
from those caused by the Van Hove singularities. For the present model, 
the density of states (per spin) is

\begin{equation}
N(\varepsilon) = \varepsilon \ \Theta(\varepsilon) \Theta(1-\varepsilon) \ + \
(2-\varepsilon) \ \Theta(\varepsilon-1) \Theta(2-\varepsilon).
\end{equation}

We use the standard Matsubara finite temperature formalism to calculate the bare and the first vertex
correction to the spin susceptibility of this model. The bare susceptibility is given by

\begin{equation}
\chi_0(\q, i \omega) = -{T \over N} \sum_{\k, i \nu} G_0 (\k, i \nu) G_0 (\k + \q, i \omega + i \nu),
\label{bare_susceptibility}
\end{equation}

\noindent where $G_0(\k, i \omega) = [i \omega - \energy (\k)]^{-1}$ 
is the bare Green's function. Doing the
frequency sum and performing analytic continuation to real frequency,
we get the standard result for the bare particle-hole susceptibility
 
\begin{equation}
\chi_0(\q, \omega) = {1 \over 4} \int d \k \
{f [ \energy(\k+\q)] - f [ \energy(\k)] \over \omega - \energy(\k + \q)
+ \energy(\k) + i \delta},
\label{baresusy}
\end{equation}

\noindent where $f(x) = [{\rm exp} \ \beta(x-\mu) + 1]^{-1}$ is the Fermi function.
At $\omega = 0$ and $\q \rightarrow {\bf 0}$ this reduces to the temperature independent Pauli susceptibility
$\chi({\bf 0},0)= N(\mu).$


\section{Susceptibility for Noninteracting Electrons}
\subsection{Imaginary Part of Susceptibility}

The susceptibility comprises real and imaginary parts
$\bare(\q, \omega) = \bare'(\q, \omega) + i \bare''(\q, \omega).$
The expression for the imaginary part is

\begin{equation}
\bare''(\q, \omega) = {\pi \over 4} \int_{-1}^1 dx \int_{-1}^1 dy \ \Big\{
{1 \over e^{\beta( \energy(x,y) - \mu )} + 1} -
{1 \over e^{\beta( \energy(x+a,y+b) - \mu )} + 1} \Big\} \
\delta \Big\{ \omega - \energy(x+a,y+b) + \energy(x,y) \Big\},
\end{equation}
where $a = {\rm max} (|q_x|,|q_y|)$ and $b = {\rm min} (|q_x|,|q_y|)$ with $0 \leq b < a \leq 1$
for $\q \in {\rm 1BZ.}$
The integral can be carried out directly by dividing the region $-1 \leq x,y \leq 1$
into 16 regions separated by the lines $x=\{ -1,-a,0,1-a,1 \}$ and $y=\{ -1,-b,0,1-b,1 \}.$
The result is conveniently expressed in terms of the following functions:

\begin{equation}
S_1(x, \omega) = {\pi \over 16} \ \Big\{ f \Big( {x- \omega \over 2} \Big)
 - f \Big( {x + \omega \over 2 } \Big) \Big\} = -S_1(x, -\omega),
\end{equation}

\begin{equation}
S_2(x) = {\pi^2 \over 24} - {\ln 2 \over 2} \Big( {x - \mu \over T} \Big)
+ {1 \over 8} \Big( {x - \mu \over T} \Big)^2
+ {1 \over 2} \ {\rm Li_2} \Big[ - \exp \Big({\mu-x \over T} \Big) \Big],
\end{equation}

\begin{equation}
S_3(x, \omega) =
\Big({\pi \omega \over 16}\Big) + \Big({\pi T \over 8} \Big)
 \ \ln \Bigg[ { \cosh \Big( {x + \omega - 2 \mu \over 4T} \Big) \over
\cosh \Big( { x - \omega - 2 \mu \over 4T} \Big)} \Bigg] \ = \
-S_3(x, -\omega),
\end{equation}

\begin{equation}
K_1 (a,b,\omega) = ( a + b - |\omega|) \ \Theta ( a + b - |\omega| ) \ \Theta( |\omega| - a + b) \ +
\ 2b \ \Theta(a - b - |\omega|),
\end{equation}

\begin{equation}
K_2 (a,b,\omega) = \Theta ( a + b - \omega) \ \Theta( \omega - a + b) \ + \
\Theta(a + b + \omega) \ \Theta(-\omega - a + b),
\end{equation}

\begin{equation}
K_3 (a,b,\omega) = \Theta ( a + b - \omega) \ \Theta( \omega + a - b) \ + \
\Theta(a + b + \omega) \ \Theta(-\omega + a - b),
\end{equation}

\begin{equation}
D_1(a,b) =
\Big( {\pi T^2 \over 2} \Big) \Big\{
S_2(1-a) - S_2(1+a) + S_2(1-b) - S_2(1+b) + S_2(2) - S_2(0) + S_2(a+b) - S_2(2-a-b) \Big\},
\end{equation}

\begin{equation}
D_2(a,b) =
\Big( {\pi T^2 \over 2} \Big) \Big\{
S_2(1+a-b) - S_2(1-a+b) + S_2(2-a) - S_2(2-b) + S_2(b) - S_2(a) \Big\}.
\end{equation}

\noindent In the definition of $S_2(x),$ we have used the dilogarithm
function which is defined by 

\begin{equation}
{\rm Li}_2(x) = \int_x^0 \ dt \ \ln(1-t)/t.
\end{equation}

\noindent Using these functions, the imaginary part of the susceptibility can be written as

\begin{equation}
\chi_0''(\q,\omega) = \reg(\q, \omega) + \sing(\q,\omega),
\end{equation}

\noindent with a regular part

\begin{equation}
\matrix{\reg(\q,\omega) \hskip-3pt &= \hskip-3pt &\Big\{S_1(4-a-b, \omega) + S_1(2 + a - b, \omega)
+ S_1(2 - a + b, \omega) + S_1(a+b, \omega) \Big\} \ K_1 (a,b, \omega ) & \cr
&+ \hskip-3pt &\Big\{ S_3 (4-a-b, \omega) - S_3(2 +a - b, \omega) 
+  S_3(2 - a + b, \omega) - S_3(a+b, \omega) \Big\} \
K_2(a,b,\omega) & \cr
&+ \hskip-3pt &\Big\{ S_3 (4-a-b, \omega)
+ S_3(2 +a - b, \omega) -  S_3(2 - a + b, \omega) - S_3(a+b, \omega) \Big\} \
K_3(a,b,\omega) & \hskip-15pt,\cr}
\end{equation}

\noindent and a singular contribution

\begin{equation}
\sing(\q, \omega)  = D_1(a,b) \ \Big\{ \delta( \omega - a - b) - \delta( \omega + a + b) \Big\} \ + \
D_2(a,b) \ \Big\{ \delta(\omega + a - b) - \delta( \omega - a + b) \Big\}.
\end{equation}

\noindent For small $(\omega/T)$ we can expand $S_1(x,\omega)$ and $S_3(x,\omega)$

\begin{equation}
S_1(x,\omega) = \Big( {\pi \omega \over 64 T} \Big) \ {\rm sech}^2 \Big( {x - 2\mu \over 4 T} \Big) +
\order ( (\omega/T)^3 ),
\end{equation}

\begin{equation}
S_3(x,\omega) = \Big( {\pi \omega \over 16} \Big) \ \Big[
1 + \tanh \Big( {x - 2 \mu \over 4T} \Big) + \order ( (\omega/T)^2 ) \Big].
\end{equation}
These expansions can be used to obtain the leading term in $(\omega/T)$ of
the regular part at the nesting vector $\Q=(\mu,\mu)$

\begin{equation}
\reg (\Q, \omega) = \Theta( 2 \mu - | \omega |) \
\Big( {\mu \pi \omega \over 32 T} \Big) \ \Big\{  1 +
2 \ {\rm sech}^2 \Big( {2 - 2 \mu \over 4 T} \Big) +
{\rm sech}^2 \Big( {4 - 4 \mu \over 4T} \Big) +
{4 T \over \mu} \ {\rm tanh} \Big( {4 - 4 \mu \over 4 T} \Big) \Big\} + \order \Big( {|\omega| \omega
\over \mu T} \Big).
\label{scaling1}
\end{equation}

\noindent For a half-filled system $(\mu = 1),$ this reduces to

\begin{equation}
\reg (\Q, \omega) = \Theta (2 - | \omega | ) \ \Big( {\pi \omega \over 8 T} \Big)
+ \order \Big( {|\omega| \omega
\over \mu T} \Big),
\end{equation}
which is just the leading term of the exact expression that 
can be obtained using the nesting condition\cite{nfl}

\begin{equation}
\reg (\Q, \omega) = (\pi / 2) \ N(1- \omega/2)
\ {\rm tanh} (\omega/ 4 T).
\label{scaling2}
\end{equation}

The momentum variation of $\reg(\q,\omega)$ is
sharply peaked at the nesting vectors as shown in \hbox{Fig. \ref{imagpart_temperature}.}
Neutron scattering experiments have observed this type of peak structure in chromium and also in
high temperature superconductors.\cite{aeppli} The strong temperature variation of the peak lineshape is
caused by nesting since the quasiparticle damping does not enter in the noninteracting susceptibility.
A 3D plot shows the full momentum dependence of $\reg(\q,\omega)$ in \hbox{Fig. \ref{imagpart_3D}.}

The imaginary part of the susceptibility at the nesting vector in 
\hbox{Eqs. (\ref{scaling1})--(\ref{scaling2})} approximately scales with frequency
divided by temperature. The weak energy dependence of the density of states, which 
is nowhere singular in this model, introduces only a slow modulation which can be neglected for
small $\omega,$ leaving only the temperature $T$ to set the effective energy scale. 
Scaling of the susceptibility was first obtained  
by means of the nesting approximation,\cite{nfl,review}
and it determines the anomalous quasiparticle damping that distinguishes the cuprates from ordinary
metals. \hbox{Fig. \ref{imagpart_frequency}} 
shows the frequency variation of $\reg(\Q,\omega)$ at several temperatures.
 
The singular part of the susceptibility comes from forward
scattering processes which are strongly enhanced in this model by the synergetic  
combination of the linear dispersion and the flatness of the Fermi surface.
These terms are key ingredients for collective charge and spin modes and they are relevant
to the issue of charge and spin separation. In \hbox{Fig. \ref{zero_sound_peak}} we
show the frequency dependence of $\chi_0''(\q,\omega)$ for small $\q.$ The delta function peak
carries a spectral weight which is two orders of magnitude larger than the integrated weight
of the continuum. Nonlinear energy dispersion or finite curvature of the Fermi surface, 
however, will remove this singularity and replace it with a very sharp, but nonetheless finite,
peak.\cite{zero_sound} The 2D nearest-neighbor tight-binding model, for example, gives a finite
susceptibility for all $\q$ except for the half-filled case when there is a Van Hove singularity
in the density of states.  


\subsection{Real Part of Susceptibility}

The real part of the susceptibility determines the SDW instability and is a
vital component in the exchange processes that may create a \dwave 
superconducting state for a system with highly anisotropic Fermi surface.
Taking the real part of \hbox{Eq. (\ref{baresusy})} we obtain the expression for the
real part of susceptibility 

\begin{equation}
\real(\q, \omega) = \Big( {1 \over 8} \Big) \ \principal \int_{-1}^1 dx \int_{-1}^1 dy \ 
{ \tanh [ \gam \ ( \energy(x,y) - \mu)] - \tanh [ \gam \ ( \energy(x+a,y+b) - \mu ) ]
\over \omega - \energy(x+a,y+b) + \energy(x,y) }. 
\end{equation}

\noindent Both integrals are principal value integrals as indicated by the symbol $\principal.$ 
The integral can be done by dividing the 
region of integration into 16 regions as was previously done for the imaginary part.    
The result can be expressed in terms of functions $R_1$ and $R_2$ which 
are defined in the Appendix. We first define several auxiliary functions:

\begin{equation}
\openup4pt \def\normalbaselines{} \matrix{ L_1(a,b) \hskip-3pt &= \ 
\ln [ \cosh \gam (2-a- \mu) ] - \ln [ \cosh \gam (2- \mu)] 
- \ln [ \cosh \gam (2-a-b-\mu)] + \ln [ \cosh \gam (2-b- \mu) \hskip1pt ] \cr 
&+ \ \ln [ \cosh \gam (1-\mu) ] \hskip1pt - \hskip2pt \ln [ \cosh \gam (1-b-\mu)]
+ \ln [ \cosh \gam (1+a-b- \mu)] - \ln [ \cosh \gam (1 + a- \mu) ] \cr
&+ \ \ln [ \cosh \gam (1-\mu) ] \hskip1pt - \hskip2pt \ln [ \cosh \gam (1+b-\mu)] 
+ \ln [ \cosh \gam (1-a+b- \mu) ] - \ln [ \cosh \gam (1-a- \mu) ] \cr
&+ \ \ln [ \cosh \gam (a-\mu)] \hskip1pt - \hskip1pt \ln [ \cosh \gam (a+b-\mu)]
+ \ln [ \cosh \gam (b- \mu)] - \ln [ \cosh \gam (- \mu)], \hfill \cr}
\end{equation}

\begin{equation}
\openup2pt \def\normalbaselines{} \matrix{ L_2(a,b, \omega) \hskip-4pt &= \hskip-3pt
&\Big( \displaystyle{b \over 16} \Big) \
\Big\{ R_1 ( 1; \ 1-a+ b; \ 2-a+b - \omega )
+ R_1 ( 1 + a -b; \ 1; \ 2+a-b - \omega ) \Big\} \cr
&+ \hskip-3pt &\Big( \displaystyle{b \over 16} \Big) \ \Big\{
R_1 ( a ; \ b; \ a+b - \omega )
+ R_1 ( 2-b; \ 2-a; \
4-a-b - \omega ) \Big\}, \hfill \cr}
\end{equation}

\begin{equation}
\matrix{ L_3(a,b, \omega) \hskip-4pt &= \hskip-3pt
&\Big( \displaystyle{a+b-\omega \over 32} \Big) \
\Big\{ R_1 ( 2-a; \ 2-a-b; \
4-a-b - \omega  )
+ R_1 ( 2; \ 2-b; \
4-a-b + \omega) \Big\} \cr
&+ \hskip-3pt &\Big( \displaystyle{a+b-\omega \over 32} \Big) \
\Big\{ R_1 (1-a+b; \ 1-a; \
2-a+b - \omega)
+ R_1 ( 1+b; \ 1; \
2-a+b + \omega) \Big\} \cr
&+ \hskip-3pt &\Big( \displaystyle{a+b-\omega \over 32} \Big) \
\Big\{ R_1 (1+a; \ 1+a-b; \
2+a-b + \omega)
+ R_1 (1; \ 1-b; \
2+a-b - \omega) \Big\} \cr
&+ \hskip-3pt &\Big( \displaystyle{a+b-\omega \over 32} \Big) \
\Big\{ R_1 (b; \ 0; \
a+b - \omega)
+ R_1 (a+b; \ a; \
a+b + \omega) \Big\}, \hfill \cr }
\end{equation}

\begin{equation}
\openup4pt \def\normalbaselines{} \matrix{ L_4(a,b, \omega)\hskip-3pt
&=\hskip-3pt &R_2 (2-a; \ 2-a-b; \ 4-a-b - \omega)
\ + \ R_2 ( 2; \ 2-b; \ 4-a-b + \omega) \cr
&+\hskip-3pt &R_2 (1-a+b; \ 1-a; \ 2-a+b - \omega)
\ + \ R_2 ( 1+b; \ 1; \ 2-a+b + \omega) \cr 
&-\hskip-3pt &R_2 (1+a; \ 1+a-b; \ 2+a-b + \omega) 
\ - \ R_2 (1; \ 1-b; \ 2+a-b - \omega) \cr
&-\hskip-3pt &R_2 (b; \ 0; \ a+b - \omega) 
\ - \ R_2 (a+b; \ a; \ a+b + \omega). \hfill \cr } 
\end{equation}

\noindent With these functions, the real part of the bare particle-hole susceptibility is

\begin{equation}
\openup2\jot \def\normalbaselines{} \matrix{ \real(\q,\omega) \hskip-3pt &= \
(T/4) \ L_1(a,b) +  
\Big( \displaystyle{D_1(a,b) \over \pi} \Big) \Big\{ \displaystyle{ 1 \over \omega + a + b} 
- \displaystyle{ 1 \over \omega - a- b} \Big\}
+ \Big( \displaystyle{D_2(a,b) \over \pi} \Big) \Big\{ \displaystyle{ 1 \over \omega - a + b} 
- {1 \over \omega + a - b} \Big\} \hfill \cr
&+ \ \Big\{L_2(a,b,\omega) \ + \ L_2(a,b,-\omega) \ + \
L_3(a,b,\omega) \ + \ L_3(a,b,-\omega) \ + \ (T/4) \ (L_4(a,b,\omega) \ + \ L_4(b,a,\omega)) \Big\}. 
\cr}
\end{equation}

The momentum variation of the real part of the susceptibility exhibits broad peaks at the nesting
vectors, as seen in \hbox{Fig. \ref{realpart_3D}} 
for a frequency $\omega = 1$ meV and bandwidth W = 2 eV. At room
temperature these nesting peaks exceed the Pauli susceptibility by roughly a factor of two. 

For $a = b = \mu = 1$ (nesting vector of a half-filled system) the expression for
the real part of the susceptibility simplifies to

\begin{equation}
\real(\Q,\omega)_{\mu=1}= - 2T \ln[\cosh(1/2T)] + \Big({2 - \omega \over 4} \Big) 
R_1(1,0,2-\omega) + \Big( {2 + \omega \over 4} \Big) R_1(1,0,2+ \omega).
\end{equation}

\noindent At the static limit and $T \rightarrow 0$ this asymptotically goes to  

\begin{equation}
\real(\Q,0)_{\mu=1} = -2T \ln [\cosh \Big( {1 \over 2 T} \Big) ] + \ln \Big( 
{2 \gamma_E \over \pi T} \Big),
\end{equation}  

\noindent where $\gamma_E = 1.781...$ is the Euler constant. 
The susceptibility at these limits can also be derived more easily
using the density of states and the nesting condition, which is exactly satisfied at
half-filled.\cite{nfl} 

The logarithmic divergence that appears in the susceptibility is vital to the 
SDW and the competing \dwave superconducting state instabilities\cite{yakovenko} 
that need to be carefully examined when we turn on the electron-electron interaction. 
In the next section we proceed to the study of vertex corrections to the susceptibility.


\section{Coulomb Interactions}

In this section we will use the standard diagrammatic analysis to discuss the 
effects of the interaction Hamiltonian $H_U$ in \hbox{Eq. (\ref{hamiltonian}).} 
The Hubbard onsite Coulomb repulsion
$U$ only couples electrons with opposite spin on the same lattice site.
This, in particular, means that all exchange diagrams vanish since
there is no interaction between electrons with equal spin. 

\subsection{Random Phase Approximation}

The RPA is one of the most useful approximation methods in many-body physics and has been used
effectively to describe plasma oscillations and density wave instabilities. In this approximation,
the longitudinal spin susceptibility for the interacting Hamiltonian is given by the simple formula

\begin{equation}
\chi^{\rm s}_{\rm rpa}(\q,iq_0)
= {2 \chi_0(\q,iq_0) \over 1 - U \chi_0(\q,iq_0)}.
\label{rpa_spin_susceptibility}
\end{equation}

\noindent The diagrams corresponding to the first three terms 
in this series are shown in \hbox{Fig. \ref{feynman_diagrams}(a)--(c)} 
where the bare susceptibility ``bubble'' is (a). Note that for each diagram there is another
diagram in which all the spins are flipped. In the paramagnetic state, which will always be assumed in
the following calculations, this spin degeneracy gives rise to the factor of two 
in \hbox{Eq. (\ref{rpa_spin_susceptibility}).} 
 
Despite its popularity, the RPA should be used with caution since
occasionally its application can lead to
serious mistakes whose remedy requires a study of vertex corrections. For example, one may
incorrectly infer that a single component electron gas should be superconducting when the
screened Coulomb interaction is treated by this method, but Sham {\it et al.} have shown that
the true situation of a net repulsion between electrons is confirmed  when vertex corrections
are taken into account.\cite{sham}

For our discussion, it is useful to define the spin-dependent particle-hole 
susceptibilities\cite{moriya}

\begin{equation}
\chi^{\sigma \sigma'} (\q,iq_0) = {1 \over N} \int_0^{\beta} d \tau 
\ e^{iq_0 \tau} \ \langle T_\tau [ n_{\sigma} (\q,\tau) 
n_{\sigma'} (-\q, 0)] \rangle,
\end{equation} 

\noindent where

\begin{equation}
n_{\sigma} (\q,0) = \sum_{\k} c_{\k+\q, \sigma}^\dagger c_{\k, \sigma}^{\vphantom{\dagger}}.
\end{equation} 

\noindent From this susceptibility we can get the charge and spin susceptibilities

\begin{equation}
\chi^{\rm c} (\q,iq_0) = \upup (\q,iq_0) + \downdown(\q,iq_0) + 
2 \updown (\q,iq_0), 
\end{equation}

\begin{equation} 
\chi^{\rm s} (\q,iq_0) = \upup (\q,iq_0) + \downdown(\q,iq_0) - 
2 \updown (\q,iq_0).
\end{equation}
 
\noindent At zeroth order in $U,$ we have $\upup (\q,iq_0) = \downdown (\q,iq_0) = \chi_0 (\q,iq_0)$
and $\updown(\q,iq_0) = 0.$ It is useful to define separate RPA series for these susceptibilities

\begin{equation}
\upup_{\rm rpa}(\q,iq_0) = {\chi_0(\q,iq_0) \over 1 - (U \chi_0(\q,iq_0))^2}
= \downdown_{\rm rpa} (\q,iq_0), 
\end{equation}

\begin{equation}
\updown_{\rm rpa}(\q,iq_0) = - {U (\chi_0(\q,iq_0))^2 \over 1 - (U \chi_0(\q,iq_0))^2}
= \downup_{\rm rpa} (\q,iq_0).
\end{equation}

\noindent Within RPA, the charge and longitudinal spin susceptibilities 
are therefore 

\begin{equation}
\chi^{\rm c}_{\rm rpa} (\q,iq_0) 
= 2(\upup_{\rm rpa} (\q,iq_0) + \updown_{\rm rpa} (\q,iq_0)),
\end{equation}

\begin{equation}
\chi^{\rm s}_{\rm rpa} (\q,iq_0) = 2(\upup_{\rm rpa} (\q,iq_0) - \updown_{\rm rpa}
(\q,iq_0)).
\end{equation}

The RPA would give an exact susceptibility if in the formula we replace the bare
susceptibility $\chi_0(\q,iq_0)$ with the exact irreducible susceptibility  
which requires all possible self energy and vertex corrections.
Clearly this is very difficult. For the tight-binding model, the spin and charge
susceptibilities have been calculated to second order in $U.$\cite{hotta} 
The calculation, however, was done on a finite lattice and the susceptibilities 
were calculated only at $\q = (0,0)$ and $\q = (\pi/a,\pi/a).$   
In the next subsection we will discuss the momentum
and temperature dependence of the leading order vertex correction to the bare 
particle-hole susceptibility for the square model.
    

\subsection{Vertex Correction to $\upup(\q,iq_0)$}

The first vertex correction to $\upup(\p,ip_0)$ is shown in \hbox{Fig. \ref{feynman_diagrams}(d)}
and is given by the expression

\begin{equation}
\vertex (\p, ip_0) = -{T \over N} \sum_{\k} \sum_{ik_0} \ \Gamma(\k,\p,ik_0,ip_0) \
G_0(\k+\p,ik_0+ip_0) \ G_0(\k,ik_0),
\label{definevertex}
\end{equation}

\noindent where 

\begin{equation}
\Gamma(\k,\p,ik_0,ip_0) = {U^2 T \over N} \sum_\q \sum_{iq_0} \chi_0 (\q,iq_0) \
G_0(\k+\q,ik_0+iq_0) \ G_0(\k+\p+\q,ik_0+ip_0+iq_0).
\end{equation}

\noindent Note that this is the only vertex correction to $\upup(\q,iq_0)$ in the second order.
To calculate this correction we define 

\begin{equation}
\vertex (\p, ip_0) = -{U^2 \over N} \sum_\k F(\k,\p,ip_0),
\end{equation}

\noindent with

\begin{equation}
F(\k,\p,ip_0) = {1 \over N} \sum_\q
{ G^{13} - G^{23} - G^{14} + G^{24} \over
(\xi_1 \ - \ \xi_2) \ (\xi_3 \ - \ \xi_4)},
\label{defineF}
\end{equation}

\begin{equation}
G^{\alpha \beta} = T^2 \ \sum_{i k_0} \ {1 \over ik_0 - \xi_\beta} \ \sum_{iq_0} 
\ {\chi_0(\q,iq_0) \over iq_0 + ik_0 - \xi_{\alpha}},
\end{equation}

\noindent where $\alpha \in \{ 1,2 \};$ $\beta \in \{3,4 \};$
$\xi_1 = \xi_{\k+\q};$ $\xi_2 = (\xi_{\k+\p+\q}- ip_0);$ $\xi_3 = \xi_\k;$
$\xi_4 = (\xi_{\k+\p} - ip_0).$ Note that $q_0$ and $p_0$ are boson frequencies, 
$k_0$ is a fermion frequency, and $\chi_0(\q,iq_0) = \real(\q,\omega \rightarrow iq_0).$
The two frequency summations in $G^{\alpha \beta}$ can be performed using 
the analytic expressions for $\chi_0(\q,iq_0)$ that we have obtained previously. 
First note that for a simple pole $1/(iq_0-z)$ we have     

\begin{equation}
\openup4pt \def\normalbaselines{} \matrix{ 
\displaystyle{ T^2 \ \sum_{ik_0} \ {1 \over ik_0 - \xi_\beta} 
\ \sum_{iq_0} \ {1 \over iq_0 + ik_0 - \xi_\alpha} \ . \ 
{1 \over iq_0 - z} \ = \ }
\cr 
\displaystyle{ [\coth(\gam z) - \tanh (\gam \xi_\alpha)] \  
\Big\{ {\tanh (\gam (\xi_\alpha - z)) - \tanh(\gam \xi_\beta) \over \xi_\alpha - z - \xi_\beta}
\Big\}.}} 
\end{equation}

\noindent Now observe that in the expressions for $\chi_0(\q,iq_0)$ 
each term has exactly one simple pole of the form 
$1/(iq_0-z_i)$ where $z_i$ varies from term to term. 
($L_1(a,b)$ and $L_3(a,b,iq_0)$ clearly are not in this form, 
but one can check easily that their sum can be written in this form.)  
Except for the poles multiplying $D_1(a,b)$ and $D_2(a,b),$ 
all the other poles reside in an integral.
We assume that the integration and the frequency summation operations commute 
so that we can perform the frequency summation first.
With all this, $G^{\alpha \beta}$ can be obtained  
from $\chi_0(\q,iq_0)$ by a simple substitution rule

\begin{equation}
\openup4pt \def\normalbaselines{} \matrix{ 
\displaystyle{{1 \over iq_0 - z_i} \ \rightarrow \  
[\coth(\gam z_i) - \tanh (\gam \xi_\alpha)]} \cr
\displaystyle{\hfill \times \ \Big\{ {\tanh (\gam (\xi_\alpha - z_i)) - 
\tanh(\gam \xi_\beta) \over \xi_\alpha - z_i - \xi_\beta}
\Big\}.}}
\end{equation}

Performing this substitution, $G^{\alpha \beta}$ 
is now expressed as a sum of one dimensional integrals. 
These integrals can be calculated analytically using the piecewise polynomial approximation to
the functions $\tanh(x),$ $\ln[\cosh(x)],$ and $\coth(x)$ as presented
in the Appendix. Note that the number of terms that need to be calculated in this method  
increases very rapidly compared to the number of terms in the bare susceptibility.
The analytic expressions, however, allow us to perform the principal value 
integrations exactly within this approximation. With this analytic expression for
$G^{\alpha \beta},$ the computation of $\vertex(\p,ip_0)$ is reduced to
a numerical computation of a four dimensional integral.

The integral over $\q$ was carried out
numerically to give the function $F(\k, \p, 0)$ defined in \hbox{Eq. (\ref{defineF}).}
We have performed these computation using SP2 parallel processors at the Maui High Performance
Computing Center and the University of Virginia. Each of the plots of $F(\k,\p,0)$ shown in 
\hbox{Fig. \ref{F_function}} typically requires about 4 days of CPU time on 10 processors.
The vertex correction $\vertex(\p,0)$ is then obtained by integrating $F(\k,\p,0)$ over $\k.$ 
 
In \hbox{Fig. \ref{vertex_momentum}} we show the result for $\vertex (\p, 0)$ at $T = 0.03$ 
at several points in the first Brillouin zone.
The influence of this vertex correction to the static susceptibility is demonstrated in
\hbox{Fig. \ref{vertex_momentum}} by the dashed curve, in comparison to the
bare susceptibility (dotted curve) and the second order RPA term (dot-dashed curve). Combining
these corrections yields the solid curve for the total susceptibility, which exceeds the 
bare curve by roughly a factor of two when $U = 0.5.$

The temperature variation of the $\vertex(\p,0)$ at the nesting vector $\Q=(\mu,\mu)$ is displayed
in \hbox{Fig. \ref{vertex_temperature}.} The bare susceptibility (solid curve) and the second order
RPA (dot-dashed curve) diverge at low temperatures as $\ln(T)$ and $\ln^3(T)$ respectively. 
The standard RPA spin susceptibility \hbox{Eq. (\ref{rpa_spin_susceptibility})} yields a N${\grave {\rm e}}$el
temperature for an SDW instability at $T_N = 60$ K for W = 2 eV and $U$ = 0.5 eV. 
The vertex correction $\vertex(\p,0)$ is expected to diverge also as $\ln^3(T)$ and we have
fitted the computed points (circles) using a third order polynomial in $\ln(T).$ 


\subsection{Vertex Corrections to $\updown(\q,iq_0)$}

At second order in $U,$ there are two vertex correction diagrams to $\updown(\q,iq_0)$
which are shown in \hbox{Fig. \ref{two_cancel}.} The expressions for these diagrams are 

\begin{equation}
\chia(q) = U^2 \sum_{k_1} \sum_{k_2}
\sum_p G_0(k_1) \ G_0(k_1+q) \ G_0(p) \ G_0(k_2 - k_1 + p)
\ G_0(k_2) \ G_0(k_2 + q),
\label{chia}
\end{equation} 

\begin{equation}
\chib(q) = U^2 \sum_{k_1} \sum_{k_2} \sum_p
G_0(k_1) \ G_0(k_1+q) \ G_0(p) \ G_0(k_2 + k_1 - p)
\ G_0(k_2) \ G_0(k_2 - q)
\label{chib}
\end{equation}

\noindent where we have used $q \equiv (\q,iq_0)$ and 

\begin{equation}
\sum_q \equiv T \sum_{iq_0} \sum_\q.
\end{equation}

\noindent These diagrams are usually expressed in terms of 
the bare particle-hole susceptibility \hbox{Eq. (\ref{bare_susceptibility})} and 
the particle-particle susceptibility  

\begin{equation}
\phi_0(\q, i \omega) = {T \over N} \sum_{\k, i \nu} G_0 (\k, i \nu) 
G_0 (\q-\k, i \omega - i \nu).
\end{equation}

\noindent We remark that at half filling these two susceptibilities are related by

\begin{equation}
\phi_0(\Q-\q,i\omega) = \chi_0(\q,i\omega),
\end{equation}

\noindent where $\Q$ is the nesting vector at half filling: $\Q = (1,1).$ For the following
discussion we will use the forms given in \hbox{Eq. (\ref{chia})-(\ref{chib}).} The fact
that the Hubbard interaction vertex does not depend on momentum directly implies  

\begin{equation}
\chia(\q,iq_0) = \vertex(\q,iq_0),
\end{equation}  

\noindent therefore our previous results for $\vertex(\q,iq_0)$ are still useful for this section.
The other diagram $\chib(\q,iq_0)$ differs from $\chia(\q,iq_0)$ only in that the particle 
lines in one of its two loops are replaced by hole propagators. In general these two diagrams
must be calculated separately. We show, however, that at half filling, where we have
exact particle-hole symmetry, these two diagrams exactly cancel each other. To show this, we write

\begin{equation}
\chia(q) + \chib(q) = U^2 \sum_{k_1} \sum_p G_0(k_1) \ G_0(k_1+q) \ G_0(p) \ \Delta(k_1,p,q),
\end{equation}

\begin{equation}
\openup4pt \def\normalbaselines{} 
\matrix{\Delta(k_1,p,q) \hskip-6pt &= \hskip-6pt &\sum_{k_2} G_0(k_2 - k_1 + p) \ G_0(k_2) \ G_0(k_2 + q) \ \cr
&+ \hskip-6pt &\sum_{k_2} G_0(k_2 + k_1 - p) \ G_0(k_2) \ G_0(k_2 - q). \cr}
\end{equation} 

\noindent The symmetry between the two terms can be seen more clearly if in the
second sum we write $k_2 = -k_3$

\begin{equation}
\openup4pt \def\normalbaselines{} 
\matrix{\Delta(k_1,p,q) \hskip-6pt &= \hskip-6pt &\sum_{k_2} 
G_0(+(k_2 - k_1 + p)) \ G_0(+k_2) \ G_0(+(k_2 + q)) \ \cr
&+ \hskip-6pt &\sum_{k_3} G_0(-(k_3 - k_1 + p)) 
\ G_0(-k_3) \ G_0(-(k_3 + q)).\cr}
\label{delta_cancellation}  
\end{equation} 

\noindent The momenta in the second sum are just the negative of the momenta that appear in the first sum.
Now consider a propagator $G(-k) = (-ik_0 - \xi(-\k))^{-1} = (-ik_0 - \xi(\k))^{-1}.$
If we perform the following transformation   

\begin{equation}
ik_0 \rightarrow ip_0; \qquad \k \rightarrow \p + \Q; \qquad \xi(\k) \rightarrow -\xi(\p); 
\end{equation}

\noindent we get $G(-k) \rightarrow (-ip_0 + \xi(\p))^{-1} 
= -G(p).$ This means that if we do the same
transformation in the second sum in \hbox{Eq. (\ref{delta_cancellation})}
by writing $ik_{30} = ik_{40}$ and $\k_3 = \k_4 + \Q,$ 
each propagator becomes the negative of the corresponding propagator in the first sum.
Since there are three propagators in each sum, this directly implies that 

\begin{equation}
\Delta(k_1,p,q) = 0.
\end{equation}

\noindent We thus conclude that $\chia(\q,iq_0) + \chib(\q,iq_0) = 0$ for a half-filled system. 
This cancellation is valid for any external momentum and frequency.
For non half-filled systems, we expect the sum to vanish in powers of ($\tildemu$/W)
where $\tildemu$ is the chemical potential measured from the center of the band, 
and W is the bandwidth. The absolute scale is determined by $\chia(\q,iq_0)$ which, 
as we have shown, is exactly equal to $\vertex(\q,iq_0)$ which has been analyzed in the 
previous section. 


\subsection{Particle-hole Symmetry}

The cancellation of diagrams presented in the previous section clearly 
can be generalized to higher order diagrams. This leads to a simple theorem: 
for a Hubbard interaction at half filling, a diagram which has a loop consisting
of $n>1$ particle lines is equal to $(-1)^n$ times a similar diagram where
the direction of all particle lines in the loop is reversed.
There are three things to note: (a) The theorem does not 
hold for the simple Hartree loop $(n = 1)$ but applies to 
any other particle loop. (b) This cancellation depends
only on the particle-hole symmetry which is exactly satisfied
at half filling. It is therefore valid for any symmetric dispersion in any dimension.
(c) In any diagram, particle lines which are not continuously 
connected to the external lines must reside in a loop. Therefore
the theorem has relevance for a large class of diagrams.

As a simple application of this theorem, consider the self-energy
diagrams in \hbox{Fig. \ref{self_energy_diagrams}.} 
The second order diagram in \hbox{Fig. \ref{self_energy_diagrams}(a)} is the first nontrivial 
self-energy diagram in a Hubbard model. The expression for this
diagram is

\begin{equation}
\Sigma(k) = U^2 \sum_q G_0(k-q) 
\chi_0(q).
\end{equation}
 
\noindent For a nested Fermi surface, Virosztek and Ruvalds have shown
that this diagram yields a quasiparticle damping which is linear in 
frequency and temperature.\cite{nfl} The RPA method, which is usually
used to selectively include higher order diagrams, replace the bare
susceptibility $\chi_0(q)$ with $\downdown_{\rm rpa}(q).$ 
The self-energy in this approximation is given by

\begin{equation}
\Sigma_{\rm rpa}(k) = \sum_{n=1}^\infty \Sigma^{(2n)}_{\rm rpa}(k),
\end{equation}

\begin{equation}
\Sigma^{(2n)}_{\rm rpa}(k) = U^{2n} \sum_q
G_0(k-q) \chi_0^{2n-1}(q).
\end{equation}

\noindent Note that there is no odd order term in this series. 
The fourth order RPA self-energy, $\Sigma^{(4)}_{\rm rpa}(k),$ is shown in 
\hbox{Fig. \ref{self_energy_diagrams}(b).} 
For a Hubbard interaction, which does not depend on momentum, this diagram is exactly
equal to another self-energy diagram due to the particle-hole ladder with four rungs, 
$\Sigma^{(4)}_{\rm ph}(k),$ which is shown in 
\hbox{Fig. \ref{self_energy_diagrams}(c).} In general, we have

\begin{equation}
\Sigma^{(2n)}_{\rm rpa}(k) = \Sigma^{(2n)}_{\rm ph}(k).
\label{rpa_and_ph}
\end{equation}

At half filling, we can use the previous theorem to relate 
the self-energy diagrams due to the particle-hole
ladder with the corresponding diagrams from the particle-particle ladder. The 
exact relation is 

\begin{equation}
\Sigma^{(m)}_{\rm ph}(k) = (-1)^m \Sigma^{(m)}_{\rm pp}(k).
\label{ph_and_pp}
\end{equation}

\noindent Combining these results, we conclude that at half filling

\begin{equation}
\Sigma^{(2n)}_{\rm rpa}(k) = 
\Sigma^{(2n)}_{\rm ph}(k) = 
\Sigma^{(2n)}_{\rm pp}(k),
\end{equation}

\begin{equation}
\Sigma^{(2n+1)}_{\rm ph}(k) + \Sigma^{(2n+1)}_{\rm pp}(k) = 0.
\end{equation} 

\noindent Using these relations, together with $\Sigma^{(2n+1)}_{\rm rpa} = 0,$ we can
write the sum of the contributions to the self-energy from the RPA, particle-hole ladder, and 
particle-particle ladder at half filling to be

\begin{equation}
\Sigma_{\rm sum}(k) = U^2 \sum_q G_0(k-q)
\chi_{\rm sum}(q),
\end{equation}

\noindent where

\begin{equation}
\chi_{\rm sum}(q) = {3 \chi_0(q) \over 1 - (U \chi_0(q))^2} - 2 \chi_0(q).
\end{equation}

\noindent We have substracted $2 \chi_0(q)$ since there is only one diagram for the
three series at the second order. 


\section{Discussion}

The present results for a square Fermi surface in general support the nested
Fermi liquid theory of the anomalous damping that distinguishes high temperature 
superconductors from ordinary metals. Analytic forms for the imaginary part of the 
susceptibility reveal a sharp peak structure in momentum space which will assure
the dominance of electron collisions across nested regions. Scaling of the susceptibility as
a function of $\omega/T$ is modified slightly by the energy dependence of the density of
states, but otherwise fits the requirements to produce a linear frequency and temperature
variation of the quasiparticle damping.

The real part of the susceptibility for the square model demonstrates a broader peak shape  
in momentum space and its maxima at the nesting vectors display a slow divergence 
in $\ln(T)$ that is important for the spin density wave   
and the competing \dwave superconducting phase instabilities. The analytic results presented 
in this paper may provide a basis for estimating the transition temperatures for Hubbard models.

The cumbersome task of evaluating the multidimensional integrals for the first vertex correction
is slightly simplified by the analytic formulas of the bare susceptibility and some approximations
that we develop for the required hyperbolic functions. The numerical evaluation of the resulting
expressions, however, still requires considerable computing power and time. The calculated 
vertex correction counteracts the susceptibility enhancement that comes from the simple RPA and
therefore should not be neglected in computations of \dwave superconducting transition temperatures.
  
Particle-hole symmetry in a half-filled energy band leads to a theorem for the cancellation
of a general class of diagrams for a Hubbard interaction.
This theorem eliminates the need to evaluate diagrams containing
loops with an odd number of propagators at half filling.

In contrast to the standard BCS theory in simple metals, where Migdal's theorem allows the
neglect of certain vertex corrections for the electron-phonon coupling, spin fluctuation theories 
normally cannot rely on a disparity between the energy scales for the interaction medium and the bandwidth.
Hence the reliability of many calculations for the \dwave 
superconducting transition temperatures are difficult to
estimate because vertex corrections for tight-binding models are particularly difficult to compute.
Our proof for cancellation of many types of higher order diagrams applies also to  
the tight-binding models at half filling and it is reasonable to expect that these diagrams will
be less important even in the realistic cases such as the copper oxide superconductors. Nevertheless
it would be interesting to derive these corrections in terms of the chemical potential 
placement away from half filling.

A quantitative constraint on our weak-coupling analysis is the strength of the Coulomb repulsion $U$ 
which should remain much less than the bandwidth. For $U/{\rm W} \ll 1,$ the combined RPA and first
vertex correction yield a net enhancement of the susceptibility. Higher order vertex correction terms
clearly should be included for larger $U.$ Self energy corrections have been examined using the nesting
approximation in the nested Fermi liquid theory.\cite{nfl} The self energy corrections for the square
model will be explored in future work. 

Finally, the controversial issue of charge and spin separation in two dimensional systems with 
nesting may benefit from the complete analytic results for the bare susceptibility which includes
the regular part in addition to the singular terms emanating from the flat sides of the Fermi surface.
Both terms should be included in future computations of the electron self energy, with particular
emphasis on particle-hole decay channels and their impact on the spectral function. Similarly these 
susceptibility formulas may feature new charge and spin collective modes which may be
relevant to a variety of interesting metals.

\bigskip
{\centerline {\bf ACKNOWLEDGMENTS}}
\bigskip

We thank V. Celli, A. Luther, C. T. Rieck, S. Tewari, H. Thacker, A. Virosztek and A. Zawadowski for
stimulating discussions. This research is 
supported by the U. S. Department of Energy Grant No. DEFG05-84ER45113. 
The numerical computations were performed using the SP2 machines at the 
Maui High Performance Computing Center and University of Virginia.


\bigskip
{\centerline {\bf APPENDIX}}
\bigskip

The functions $R_1$ and $R_2$ are defined by (the dependence on $\gam$ and $\mu$ are implicit)

\begin{equation}
R_i (a;b;c) =
M_i (\gam (a-\mu), \gam(c/2 - \mu)) - 
M_i (\gam (b-\mu), \gam (c/2 - \mu)), \ \ i \in \{ 1,2 \},
\end{equation}
with

\begin{equation}
M_1 (x, w) = \principal \int_0^x {{\rm tanh} (z) \over z - w} \ dz,
\end{equation}

\begin{equation}
M_2 (x, w) = \principal \int_0^x { \ln \cosh (z) \over z - w} \ dz.
\end{equation}
These integrals cannot be expressed in terms of known functions. In this paper
we have chosen to analyze these integrals by approximating the functions $\tanh (x)$ and
$\ln [\cosh(x)]$

\begin{equation}
\tanh (x) \rightarrow x \Theta(c_1 - |x|) + \Big\{ [ \ 1 + m (|x| - c_2)^3 \ ] \
\Theta(|x|-c_1) \Theta(c_2 - |x|) + \Theta(|x| - c_2) \Big\} \ {\rm sign}(x),
\end{equation}

\begin{equation}
\ln [\cosh (x)] \rightarrow {x^2 \over 2} \Theta(c_1 - |x|) + \{ |x| + {m \over 4} (|x| - c_2)^4 - c_3 \} \
\Theta(|x|-c_1) \Theta(c_2 - |x|) + \{ |x| - c_3 \} \Theta(|x| - c_2),
\end{equation}
with $c_1 = (4/25);$ $m = (625/11907);$ $c_2 = (67/25);$ $c_3 = (1691/2500).$
Using these approximations, the integrals can be analytically calculated. For $M_1(x,w)$ we obtain

\begin{equation}
M_1 (x, w) =
Q(x, w) \ \Theta(x) - Q(-x, -w) \ \Theta(-x),
\end{equation}

\begin{equation}
Q(x,w) = Q_1(x, w) \ \Theta (c_1 - x) \ + \
Q_2 (x, w) \ \Theta(x - c_1) \Theta(c_2 - x) \ + \
Q_3 (x, w) \ \Theta(x - c_2),
\end{equation}

\begin{equation}
Q_0(x,w) = \sum_{n=1}^3 \displaystyle{1 \over n} (w-c_2)^{3-n} (x-c_2)^n
+ (w-c_2)^3 \ \ln |x - w|,
\end{equation}

\begin{equation}
Q_1(x,w) = x + w \ \ln \Big| { x-w \over w} \Big|,
\end{equation}

\begin{equation}
Q_2(x,w) = Q_1(c_1, w) +
m \ (Q_0(x,w) -  Q_0(c_1,w)) + \ln \Big| {x - w \over c_1 - w} \Big|,
\end{equation}

\begin{equation}
Q_3(x,w) = Q_2(c_2, w) + \ln \Big| {x-w \over c_2 - w} \Big|.
\end{equation}
Similarly, for $M_2(x,w)$ we get

\begin{equation}
M_2 (x, w) =
B(x, w) \ \Theta(x) + B(-x, -w) \ \Theta(-x),
\end{equation}

\begin{equation}
B(x,w) = B_1(x, w) \ \Theta (c_1 - x) \ + \
B_2 (x, w) \ \Theta(x - c_1) \Theta(c_2 - x) \ + \
B_3 (x, w) \ \Theta(x - c_2),
\end{equation}

\begin{equation}
B_0(x,w) = \sum_{n=1}^4 \displaystyle{1 \over n} (w-c_2)^{4-n} (x-c_2)^n
+ (w-c_2)^4 \ \ln |x - w|,
\end{equation}

\begin{equation}
B_1 (x, w)
= {x^2 \over 4}  + {w x \over 2} + {w^2 \over 2} \ \ln \Big| { x-w \over w} \Big|,
\end{equation}

\begin{equation}
B_2 (x, w)
= B_1(c_1,w) + (m/4) \ (B_0(x,w) - B_0(c_1,w)) + (x - c_1) + (w-c_3) \ \ln \Big|
{x - w \over c_1 - w} \Big|,
\end{equation}

\begin{equation}
B_3 (x, w) = B_2(c_2,w) + (x-c_2) + (w-c_3) \ln \Big| {x - w \over c_2 - w} \Big|.
\end{equation}

Finally, in the calculation of the vertex correction, we also need to approximate the
function $\coth(x).$ We do this using the following polynomial

\begin{equation}
\coth(x) \rightarrow (1/x) \Theta(c_4 - |x|) + {\rm sign}(x) (1 + c_5/x^2) \Theta(|x| - c_4),
\end{equation}

\noindent with $c_4 = 1/2$ and $c_5 = 1/4.$

\begin{figure}
\caption{Fermi surface for the square model using a chemical potential $\mu = 0.9.$
The arrows denote the nesting vectors for this Fermi surface.}
\label{square_fermi_surface}
\end{figure} 

\begin{figure}
\caption{$\reg(\q,\omega)$
shows sharp peaks at the nesting vectors
$\Q_1=(\mu,\mu)$ and $\Q_2=(2-\mu,2-\mu)$ at low temperatures.
The model parameters are chosen to be W = 2 eV, $\omega = 1$ meV, and $\mu = 0.9$ eV.
The three curves are for 2$T$/W = 0.01, 0.02, and 0.03.
These nesting peaks contrast with isotropic electronic structure results
which yield a susceptibility that has a smooth momentum dependence.}
\label{imagpart_temperature}
\end{figure}

\begin{figure}
\caption{The momentum dependence of $\reg(\k,\omega)$ for $\mu = 0.9,$ $\omega = 0.01$
and $T = 0.03.$ The peaks at the nesting vectors are strongly temperature dependent.}
\label{imagpart_3D}
\end{figure}

\begin{figure}
\caption{Frequency variation of $\reg(\Q,\omega)$
is shown at 60 K (solid curve), 120 K (dashed curve), and 240 K (dotted curve)
using W = 2 eV and $\mu = 0.9$ eV.}
\label{imagpart_frequency}
\end{figure}

\begin{figure}
\caption{Frequency dependence of $\chi_0''(\q,\omega)$ for $\q = (q,q)$ consists of
a regular particle-hole continuum and a singular peak. The peak has the form 
$A(q) \delta(\omega-2q)$ with $A(q)$ increases linearly with $q$ as shown in the inset.}
\label{zero_sound_peak}
\end{figure}

\begin{figure}
\caption{Real part of the susceptibility $\chi'(\k,\omega)$ 
in the static limit ($\omega = 0$) as a function
of momentum shows the nesting peaks. We use $T = 0.03$ and $\mu = 0.9.$ Note that
the vertical scale begins from 0.9, which is the density of states at the chemical 
potential $N(\mu).$}
\label{realpart_3D}
\end{figure}

\begin{figure}
\caption{The first three terms of the RPA series (a--c)         
and the second order vertex correction diagram (d) for $\chi^{\uparrow \uparrow}.$ 
Note that diagram (b) belongs to $\chi^{\uparrow \downarrow}.$}
\label{feynman_diagrams}
\end{figure}

\begin{figure}
\caption{Clockwise from top left: the function $F(\k, \p, 0)$ plotted as a function of $\k$
for $\p$ = (0.1, 0), (0.5, 0), (0.9, 0.9), and (0.9, 0).}
\label{F_function}
\end{figure}

\begin{figure}
\caption{The static susceptibility as a function of momentum showing
the bare susceptibility $\chi(\q,0)$ (dotted curve), second order
RPA contribution $U^2 {\chi}^3(\q,0)$ (dashed curve), vertex correction
$\chi_{\rm corr}(\q, 0),$ and the total of these three terms (solid curve).
$U$ = 0.5 (W/2).}
\label{vertex_momentum}
\end{figure}

\begin{figure}
\caption{The static susceptibility at a nesting vector
$\chi(\Q,0)$ diverges as ln($T$) at low temperatures as shown by the
solid curve. The second order RPA contribution $U^2 {\chi}^3(\Q,0),$
which diverges as $\ln^3(T),$ is shown by the dot-dashed curve.
The circles represent the results of our numerical computation
for $\chi_{\rm corr}(\Q,0).$ This vertex correction is expected to
diverge as $\ln^3(T)$ and we have drawn a fit (dotted curve) for the data
using a polynomial $3.766 + 3.483 x + 0.9530 x^2 + 0.08844 x^3$ where
$x$ = ln(2$T$/W). We have used $\mu = 0.9$ (W/2) and $U$ = 0.5 (W/2).}
\label{vertex_temperature}
\end{figure}

\begin{figure}
\caption{The first two vertex correction diagrams for $\updown(\q,iq_0).$ 
The diagrams only differ in the orientation of lines in the second loop. 
These two diagrams exactly cancel each other at half filling.}
\label{two_cancel}
\end{figure}

\begin{figure}
\caption{Diagrams (a) and (b) are the first two self energy diagrams
in the RPA; diagram (c) is the third term in the self energy series due to
the particle-hole ladder; diagram (d) is the third term in the 
particle-particle ladder. For a Hubbard onsite interaction diagrams
(b) and (c) are equal and at half filling diagram (d) is also equal to 
diagram (c).}  
\label{self_energy_diagrams}
\end{figure}


\begin{references}
\bibitem{overhauser} A. Overhauser, Phys. Rev. {\bf 128}, 1437 (1962).

\bibitem{gruner} See, for example, G. Gr$\ddot{\rm u}$ner, {\it Density Waves in Solids}, 
Addison-Wesley, Reading, 1994.

\bibitem{nfl} A. Virosztek and J. Ruvalds, Phys. Rev. B {\bf 42}, 4064 (1990).

\bibitem{review} J. Ruvalds, Superconductor Science and Technology {\bf 9}, 905
(1996).

\bibitem{shen} Z. X. Shen and D. S. Dessau, Phys. Rep. {\bf 253}, 1 (1995).

\bibitem{berk} N. F. Berk and J. R. Schrieffer, Phys. Rev. Lett. {\bf 17}, 433 (1966).

\bibitem{scalapino} D. J. Scalapino, E. Loh, Jr. and J. E. Hirsch, Phys. Rev. B {\bf 35}, 
6694 (1987); D. J. Scalapino, Phys. Rep. {\bf 250}, 329 (1995).

\bibitem{dpair} J. Ruvalds, C. T. Rieck, S. Tewari, J. Thoma and A. Virosztek,\
Phys. Rev. B {\bf 51}, 3797 (1995).

\bibitem{schrieffer} J. R. Schrieffer, J. Low Temp. Phys. {\bf 99}, 397 (1995).

\bibitem{mahan} G. Mahan, {\it Many-Particle Physics}, Plenum, NY, 1990.

\bibitem{vertexpreprint} A. Virosztek and J. Ruvalds, preprint.

\bibitem{mattis} D. C. Mattis, Phys. Rev. B {\bf 36}, 745 (1987).

\bibitem{luther} A. Luther, Phys. Rev. B {\bf 50}, 11446 (1994).

\bibitem{hlubina} R. Hlubina, Phys. Rev. B {\bf 50}, 8252 (1994).

\bibitem{manybody} S. Tomonaga, Prog. Theor. Phys. (Kyoto) {\bf 5}, 544 (1950); 
D. C. Mattis and E. H. Lieb, J. Math. Phys. {\bf 6}, 304 (1965); A. Luther and 
I. Peschel, Phys. Rev. B {\bf 12}, 3908 (1975).
  
\bibitem{li} Y. M. Li, Phys. Rev. B {\bf 51}, 13046 (1995).

\bibitem{haldane} F. D. M. Haldane, J. Phys. C {\bf 14}, 2585 (1981).

\bibitem{anderson} P. W. Anderson, Physics Today {\bf 50}(10), 42 (1997).

\bibitem{aeppli} See, for example, G. Aeppli {\it et al.}, in {\it Perspectives in
Many-Particle Physics}, Proceedings of the International School of Physics
``Enrico Fermi'', Course 121, Varenna, 1992, edited by R. A. Broglia, J. R. Schrieffer
and P. F. Bortignon, North-Holland, New York, 1994.

\bibitem{zero_sound} D. Djajaputra and J. Ruvalds, in preparation.

\bibitem{yakovenko} A. T. Zheleznyak, V. M. Yakovenko and I. E. Dzyaloshinskii,
Phys. Rev. B {\bf 55}, 3200 (1997).

\bibitem{sham} H. Rietschel and L. J. Sham, Phys. Rev. B {\bf 28}, 5100 (1983);
M. Grabowski and L. J. Sham, Phys. Rev. B {\bf 29}, 6132 (1984). 

\bibitem{moriya} T. Moriya, {\it Spin Fluctuations in Itinerant Electron Magnetism},
Springer-Verlag, Berlin, 1985.

\bibitem{hotta} T. Hotta and S. Fujimoto, Phys. Rev. B {\bf 54}, 5381 (1996).

\end{references}
\end{document}